\documentclass[cits]{PoS}

\title{QCD Results from the Fermilab Tevatron $p\bar{p}\,$ Collider}

\ShortTitle{QCD Tevatron Results}

\author{\speaker{Andreas Warburton}\thanks{Supported by the Natural Sciences and Engineering Research Council of Canada.} \\ { (on behalf of the CDF and D\O \ Collaborations)}\\ \\
        Rutherford Physics Building, McGill University\\ Montr\'eal, Qu\'ebec~~H3A~2T8\\ Canada\\
        E-mail: \email{andreas.warburton@mcgill.ca}}

\abstract{
Selected recent quantum chromodynamics (QCD) measurements are reviewed for Fermilab Run~II Tevatron proton-antiproton collisions studied by the Collider Detector at Fermilab (CDF) and D\O \ Collaborations at a centre-of-mass energy of $\sqrt{s}=1.96$~TeV.  Tantamount to Rutherford scattering studies at the TeV scale, inclusive jet and dijet production cross-section measurements are used to seek and constrain new particle physics phenomena, test perturbative QCD calculations, inform parton distribution function (PDF) determinations, and extract a precise value of the strong coupling constant, $\alpha_s(m_Z) = 0.1161^{+0.0041}_{-0.0048}$.  Inclusive photon production cross-section measurements reveal an inability of next-to-leading-order (NLO) perturbative QCD (pQCD) calculations to describe low-energy photons arising directly in the hard scatter.  Events with $\gamma + 3$-jet configurations are used to measure the increasingly important double parton scattering (DPS) phenomenon, with an obtained effective interaction cross section of $\sigma_{\rm eff} = 16.4 \pm 2.3$~mb.  Observations of central exclusive particle production demonstrate the viability of observing the Standard Model Higgs boson using similar techniques at the Large Hadron Collider (LHC).  Three areas of inquiry into lower energy QCD, crucial to understanding high-energy collider phenomena, are discussed: the examination of intra-jet track kinematics to infer that jet formation is dominated by pQCD, and not hadronization, effects; detailed studies of the underlying event and its universality; and inclusive minimum-bias charged-particle momentum and multiplicity measurements, which are shown to challenge the Monte Carlo generators.
}

\FullConference{XXth Hadron Collider Physics Symposium \\
		 November 16 -- 20, 2009 \\
		 Evian, France}

\begin{document}

\section{Introduction}
\vspace*{-0.3cm}
Quantum chromodynamics (QCD), the theory of the strong interaction between quarks and gluons, is intrinsic to experimental studies of hadron collisions.  This paper reviews several recent QCD results from the CDF and D\O \ experiments in analyses of $\sqrt{s}=1.96$~TeV $p\bar{p}$ collisions covering up to 2.7~fb$^{-1}$ of time-integrated luminosity.  The general approach has been to search for new physics phenomena, test QCD theory, enable electroweak and exotic measurements by informing Monte Carlo (MC) background models, and lay the groundwork for the LHC era of $pp$ collisions.

\section{Inclusive Jet and Dijet Production}
\begin{figure}[h]
\vspace*{-0.5cm}
\includegraphics[width=.35\textwidth]{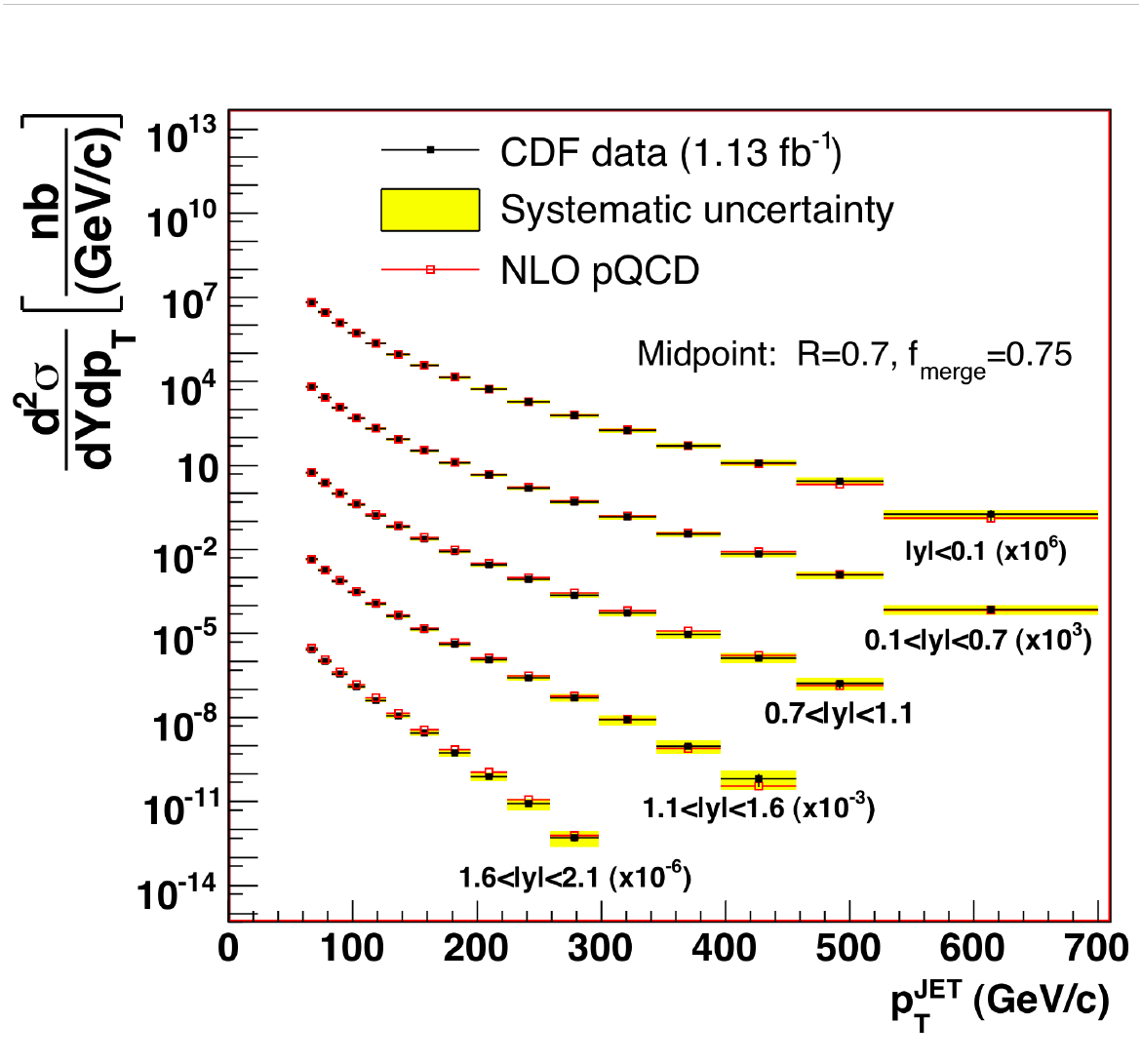}
\includegraphics[width=.31\textwidth]{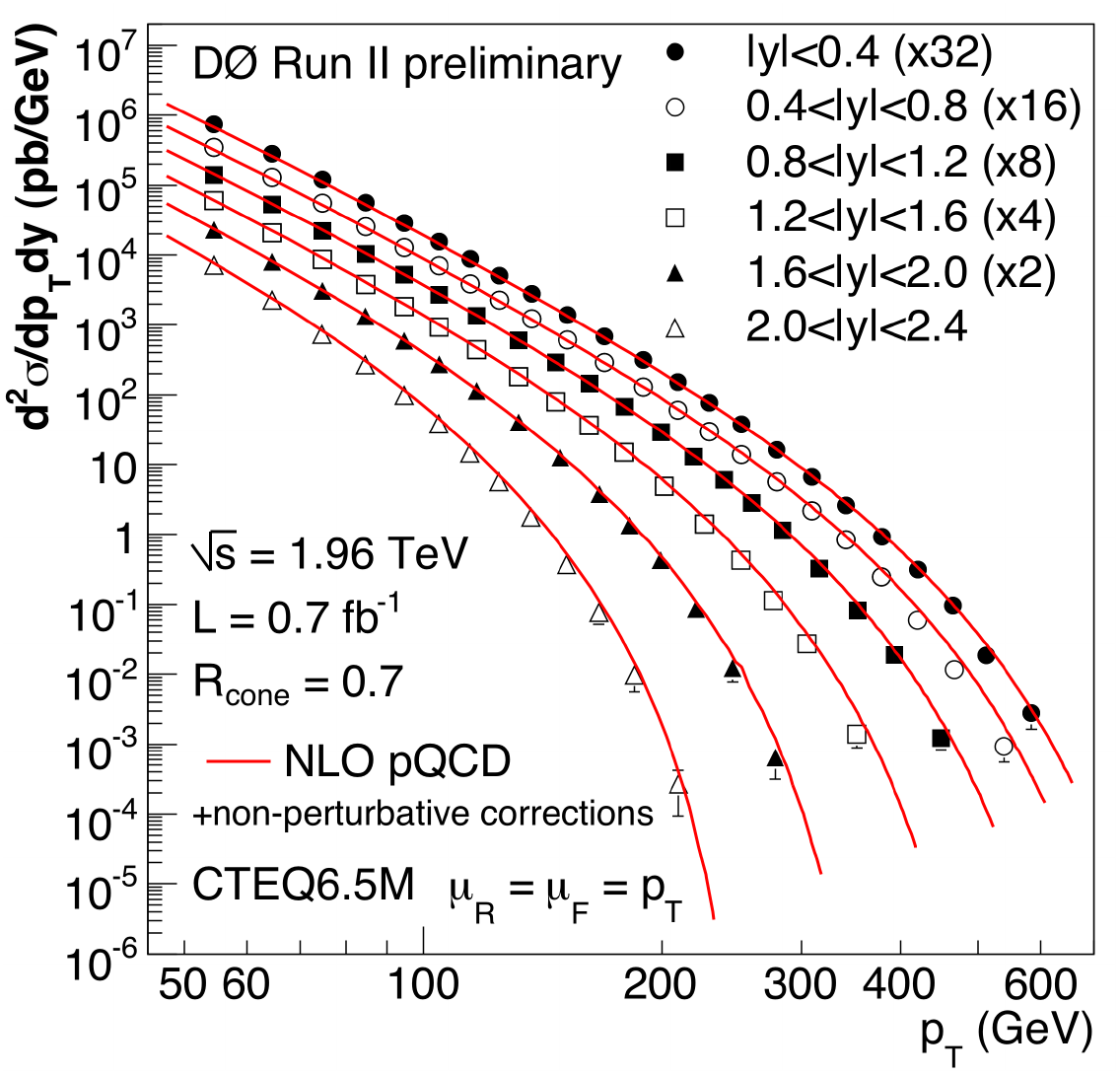}
\includegraphics[width=.295\textwidth]{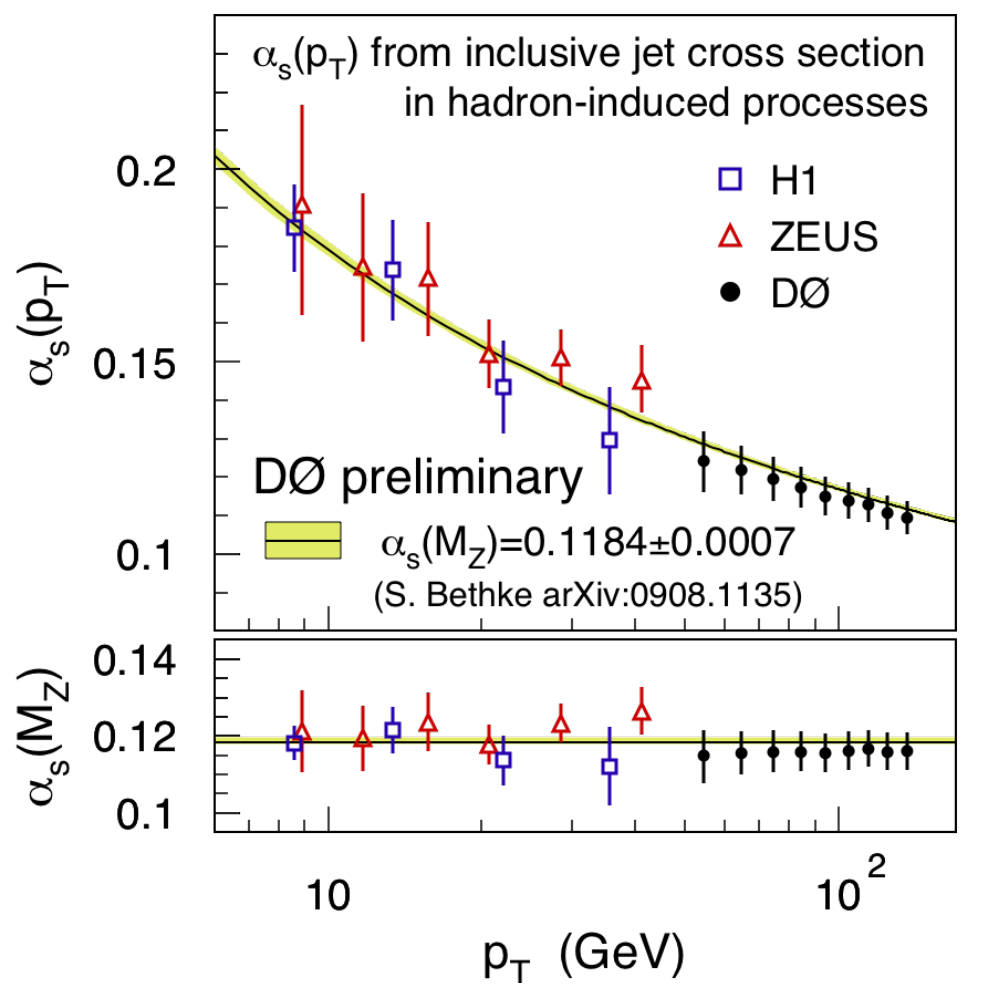}
\caption{[left] CDF~\cite{CDF:InclJet} and [centre] D\O \ \cite{D0:InclJet} inclusive jet differential
cross sections, for multiple rapidity regions, compared to NLO QCD; [right] derived $\alpha_s(p_T)$ and $\alpha_s(m_Z)$ values as a function of jet $p_T$~\cite{D0:alpha_s}.}
\label{fig:InclJet}
\end{figure}

\vspace*{-0.3cm}
Heedful testing of perturbative QCD (pQCD) at the shortest distance scales ever studied in collider experiments is provided through measurements of differential inclusive jet and dijet cross sections.  Recent CDF~\cite{CDF:InclJet} and D\O \ \cite{D0:InclJet} results, which agree mutually and with next-to-leading-order (NLO) pQCD calculations over eight orders of magnitude, are shown in Fig.~\ref{fig:InclJet}.  Articulation of the Tevatron measurements into several rapidity regions has revealed that the experimental precision now exceeds that of the parton distribution function (PDF) uncertainties and that softer gluons at high Feynman $x$ ($\geq$0.25) need to be included in the PDF fits.

Also shown in Fig.~\ref{fig:InclJet} are recently extracted correlated values of the strong coupling constant $\alpha_s$ by D\O \ \cite{D0:alpha_s}.  Inclusive cross sections for jets with transverse momentum $50 < p_T < 145$~GeV, corresponding to $x\leq 0.25$, were used to ensure negligible correlations between PDFs and extracted $\alpha_s$ values and to derive a mean of $\alpha_s(m_Z) = 0.1161^{+0.0041}_{-0.0048}$~\cite{D0:alpha_s}.  The precise D\O \ measurements lie at energies intermediate to results from the HERA experiments (Fig.~\ref{fig:InclJet}) and to those from an earlier CDF Run~I result~\cite{CDF:alpha_s}, which spanned $40 < p_T < 250$~GeV.

\begin{figure}
\vspace*{-0.5cm}
\includegraphics[width=.36\textwidth]{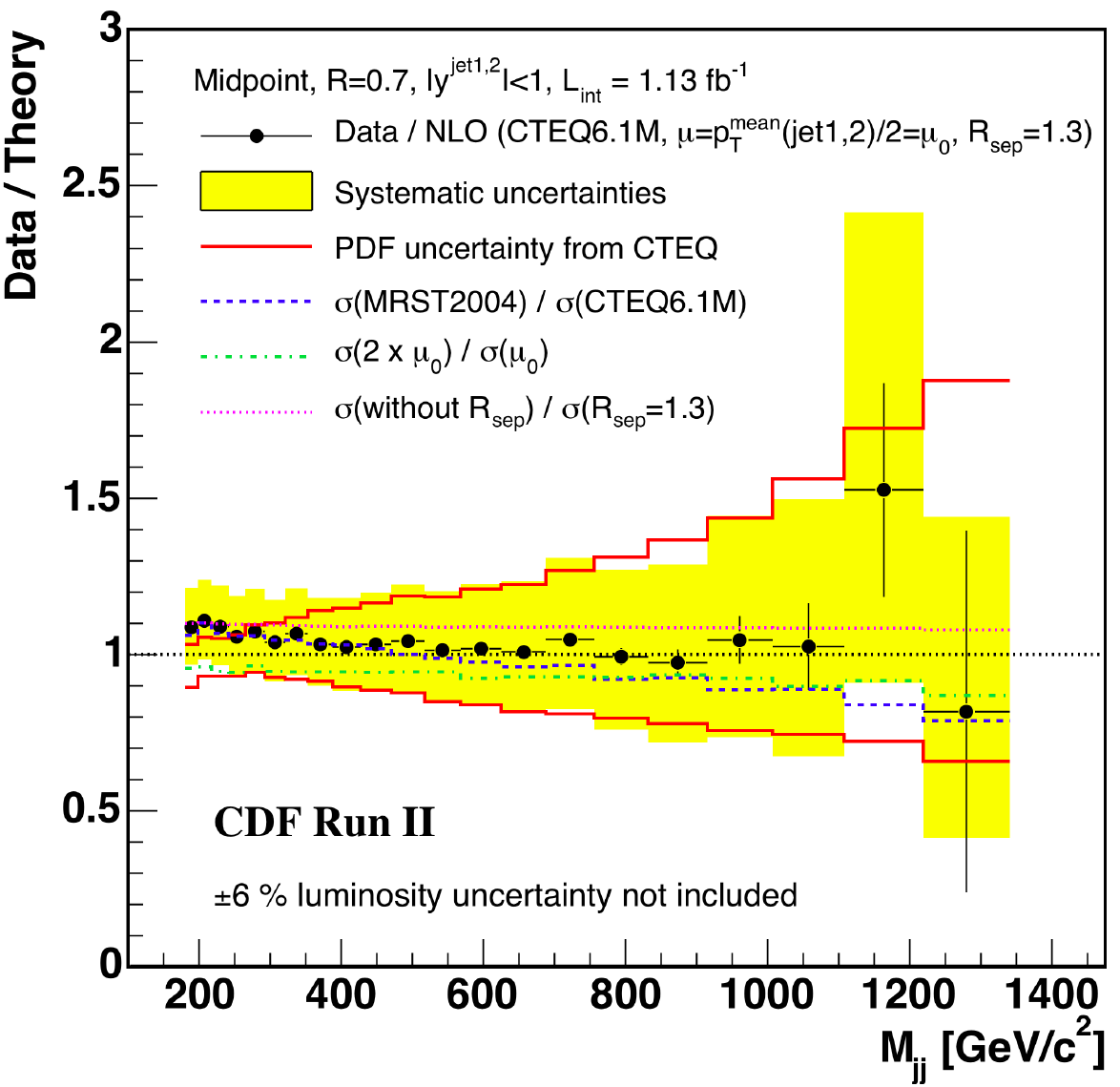}
\includegraphics[width=.35\textwidth]{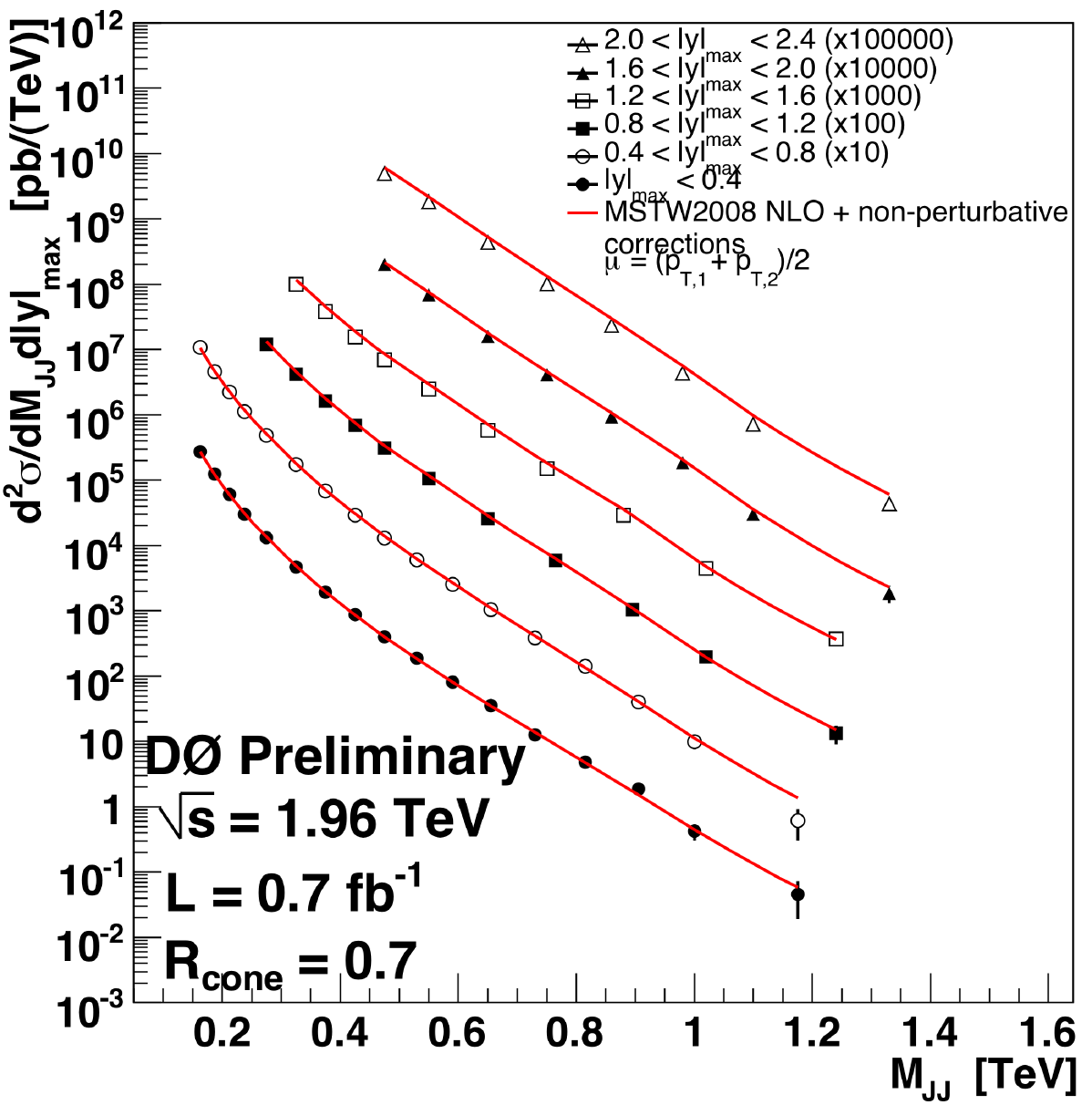}
\includegraphics[width=.28\textwidth]{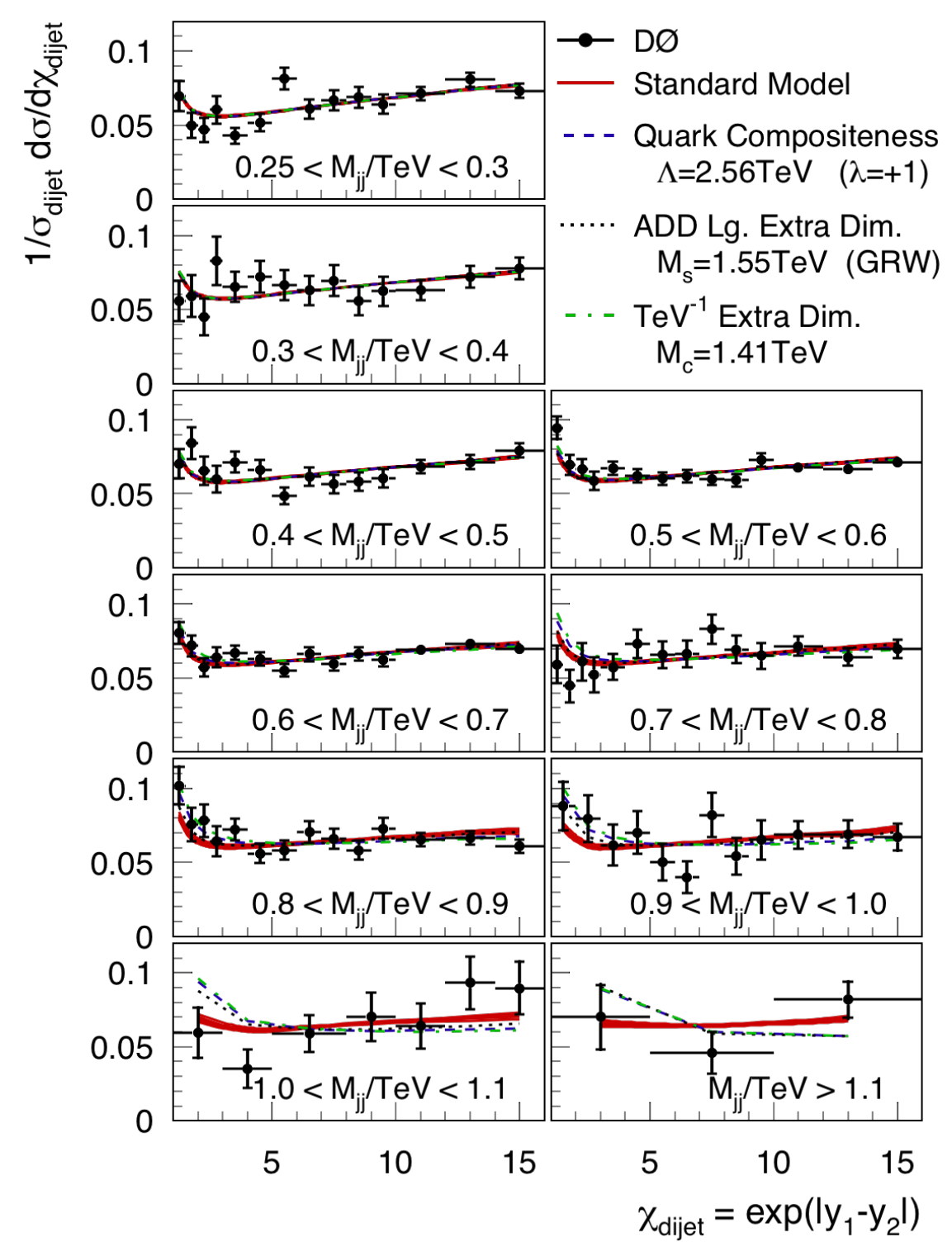}
\caption{[left] CDF~\cite{CDF:dijet_mass} and [centre] D\O \ \cite{D0:dijet_mass} dijet-mass differential cross sections compared to NLO QCD predictions; [right] D\O \ dijet angular distributions~\cite{D0:dijet_angular}.}
\label{fig:Mjj}
\end{figure}

Inclusive dijet observables are favoured harbingers of new physics phenomena.  Fig.~\ref{fig:Mjj} shows recent results from CDF~\cite{CDF:dijet_mass} and D\O \ \cite{D0:dijet_mass} of dijet invariant mass ($M_{jj}$) distributions, for which Standard Model (SM) extensions predict resonant enhancements.  Both experiments demonstrate a mass reach out beyond 1.2~TeV/$c^2$ and test pQCD predictions, with no indications of resonances.  CDF sets the tightest 95\% confidence level (CL) mass exclusion limits to date on several exotic particle species that decay to two jets: excited quarks, colour-octet $\rho_T$ mesons, axigluons, flavour-universal colorons, and $W^\prime$ and $Z^\prime$ bosons~\cite{CDF:dijet_mass}.  D\O \ demonstrates PDF sensitivity at forward maximum rapidities in the range $2.0 < \left| y_{\rm max}\right| < 2.4$, showing softer high-$x$ gluons to be favoured~\cite{D0:dijet_mass}.

Fig.~\ref{fig:Mjj} also shows consistency between pQCD and a D\O \ measurement of the dijet angular distribution, an observable that is complementary to $M_{jj}$ in its capacity to reveal new phenomena~\cite{D0:dijet_angular}.  Rutherford and QCD scattering are largely insensitive to $\chi_{\rm dijet}$, which is expected to peak at low values (central rapidities) for new physics such as composite quarks, large extra spatial dimensions (LEDs), and TeV$^{-1}$ scale extra dimensions.  D\O \ has set the most stringent 95\% CL limits on the quark compositeness scale $\Lambda > 2.9$~TeV, the ADD LED (GRW) effective Planck scale $M_S > 1.6$~TeV, and the TeV$^{-1}$ ED compactification scale $M_C > 1.6$~TeV~\cite{D0:dijet_angular}.

\section{Inclusive Photon Production and Double Parton Scattering}
\begin{figure}
\includegraphics[width=.54\textwidth]{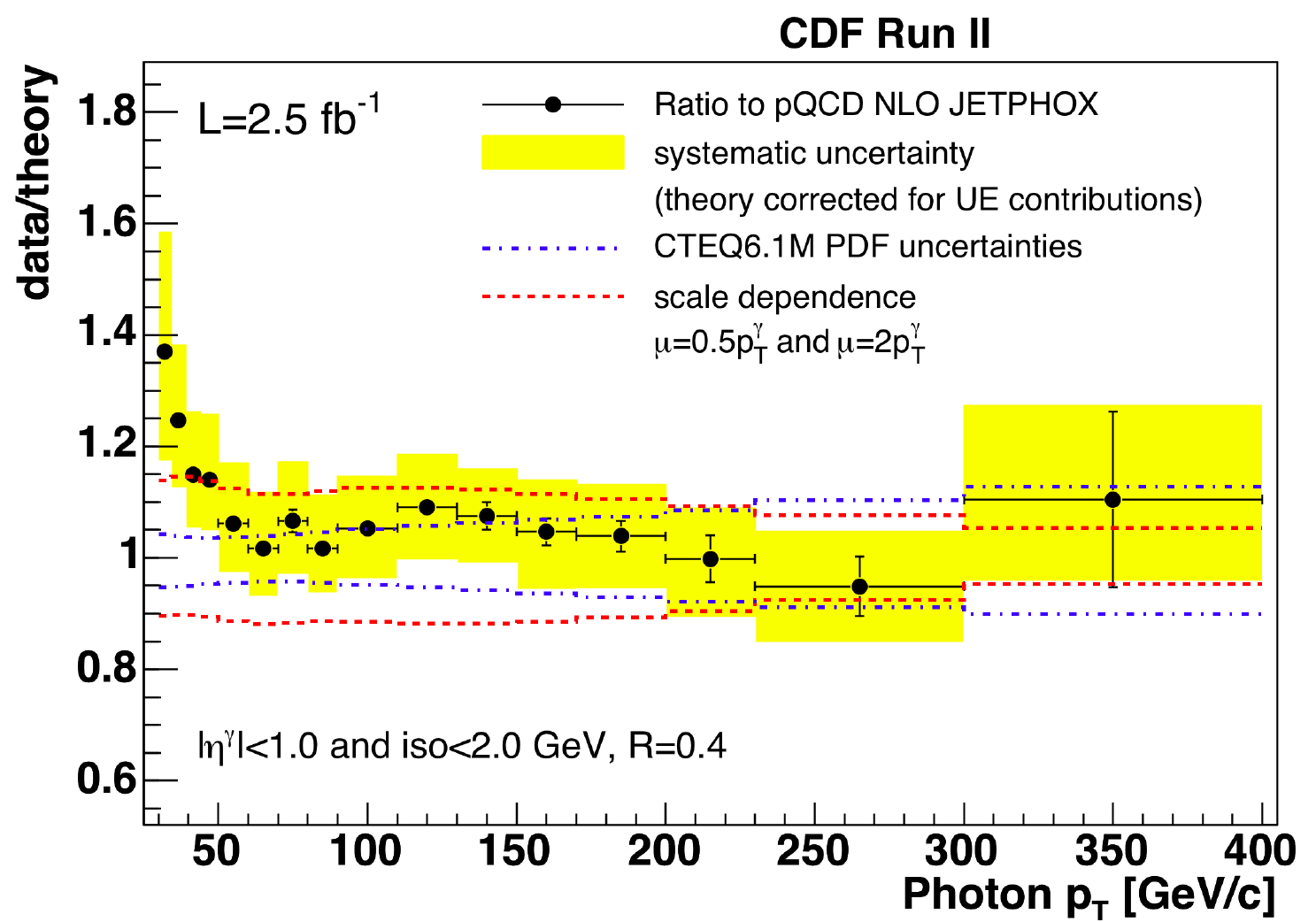}
\includegraphics[width=.41\textwidth]{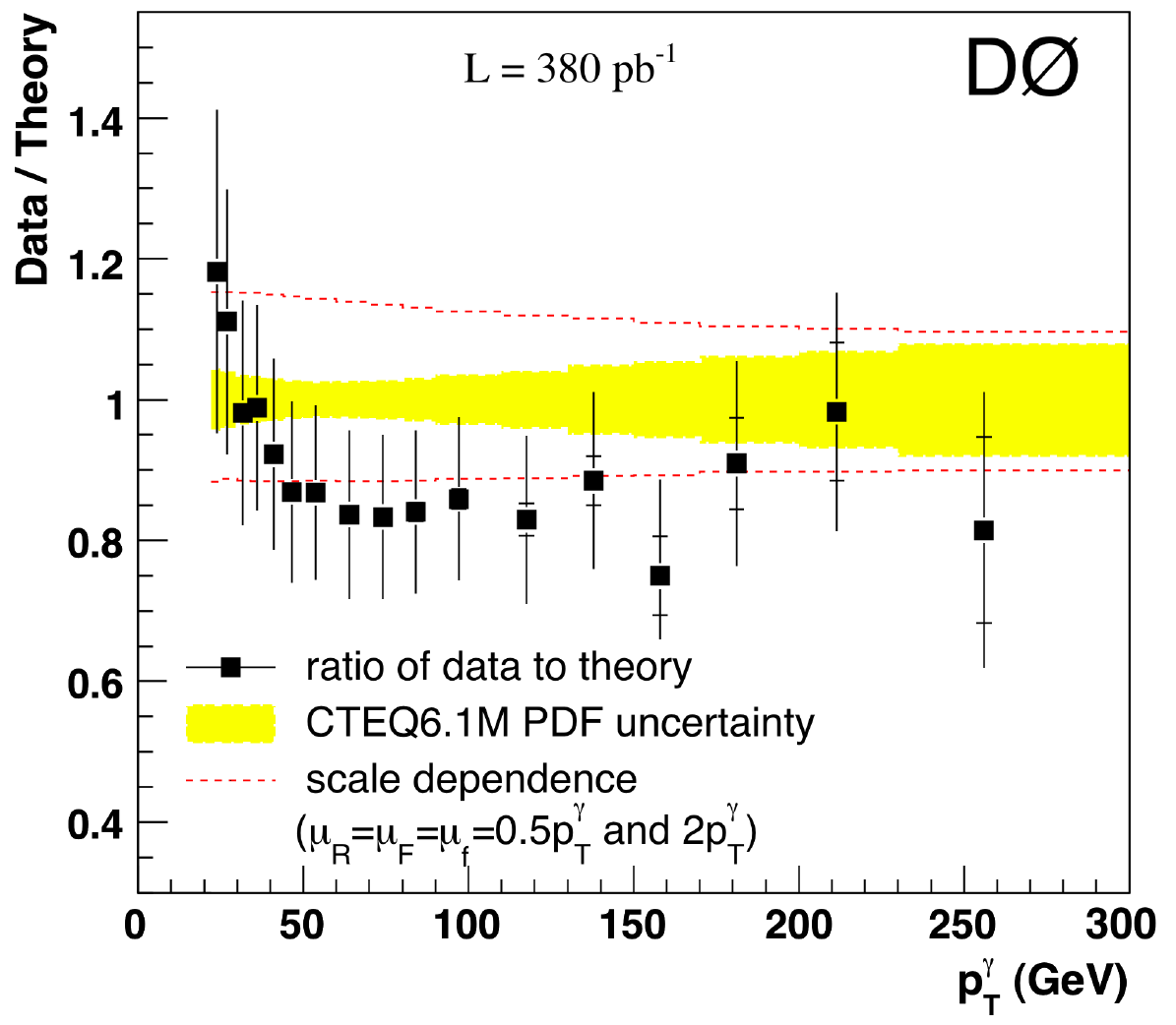}
\caption{[left] CDF~\cite{CDF:photon} and [right] D\O \ \cite{D0:photon} inclusive photon cross sections compared to NLO QCD.}
\label{fig:photon}
\end{figure}

\vspace*{-0.3cm}
Photons emerging directly from $p\bar{p}$ collisions are unaffected by fragmentation and hadronization, and therefore serve as potent probes of the hard parton scattering dynamics, with potential sensitivity to gluon PDFs.  Fig.~\ref{fig:photon} depicts recent CDF~\cite{CDF:photon} and D\O \ \cite{D0:photon} inclusive photon production cross-section measurements, which agree within experimental uncertainties and show that NLO pQCD has difficulty describing the behaviour in data at low photon $p_T$.  Further theoretical effort will be necessary to understand this discrepant region.

D\O \ has studied $\gamma + 3$-jet events to measure double parton scattering (DPS), whereby two pairs of partons undergo hard interactions in a single $p\bar{p}$ collision.  DPS is not only a background to many rare processes, especially at higher luminosities, but also provides insight into the spatial distribution of partons in the colliding hadrons.  The DPS cross section is expressed as $\sigma^{\gamma +3 {\rm jet}}_{\rm DPS}\equiv\sigma_{\gamma j}\sigma_{jj}/\sigma_{\rm eff}$, where $\sigma_{\rm eff}$ is the effective interaction region that decreases for less uniform spatial parton distributions.  D\O \ measures a mean of $\sigma_{\rm eff} = 16.4 \pm 2.3$~mb~\cite{D0:DPS}, which is consistent with an earlier CDF result~\cite{CDF:DPS}, and finds $\sigma_{\rm eff}$ to be independent of jet $p_T$ in the second interaction.  If more precise studies can reveal a $\sigma_{\rm eff}$ sensitivity to jet $p_T$, this could indicate a dynamical departure from the na\"ive assumption that $\sigma^{\gamma +3 {\rm jet}}_{\rm DPS}$ depends on an uncorrelated product of PDFs in $\sigma_{\gamma j}$ and $\sigma_{jj}$~\cite{Snigirev}.

\section{Central Exclusive Particle Production}
\begin{figure}
\includegraphics[width=.47\textwidth]{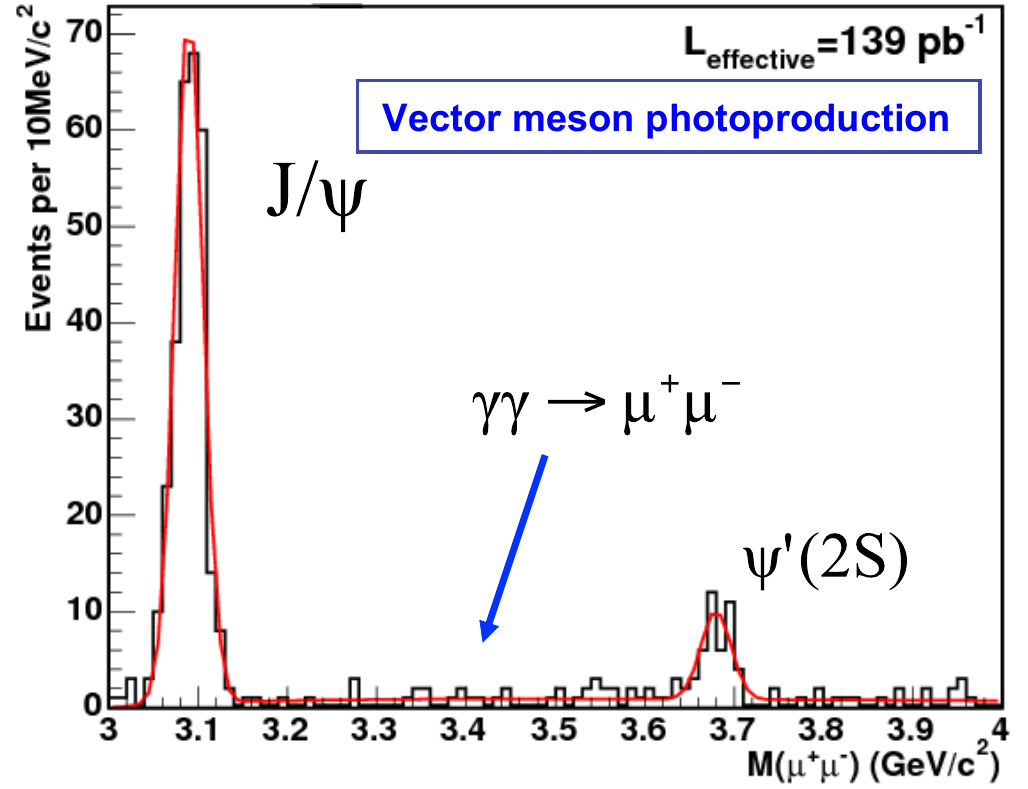}
\includegraphics[width=.40\textwidth]{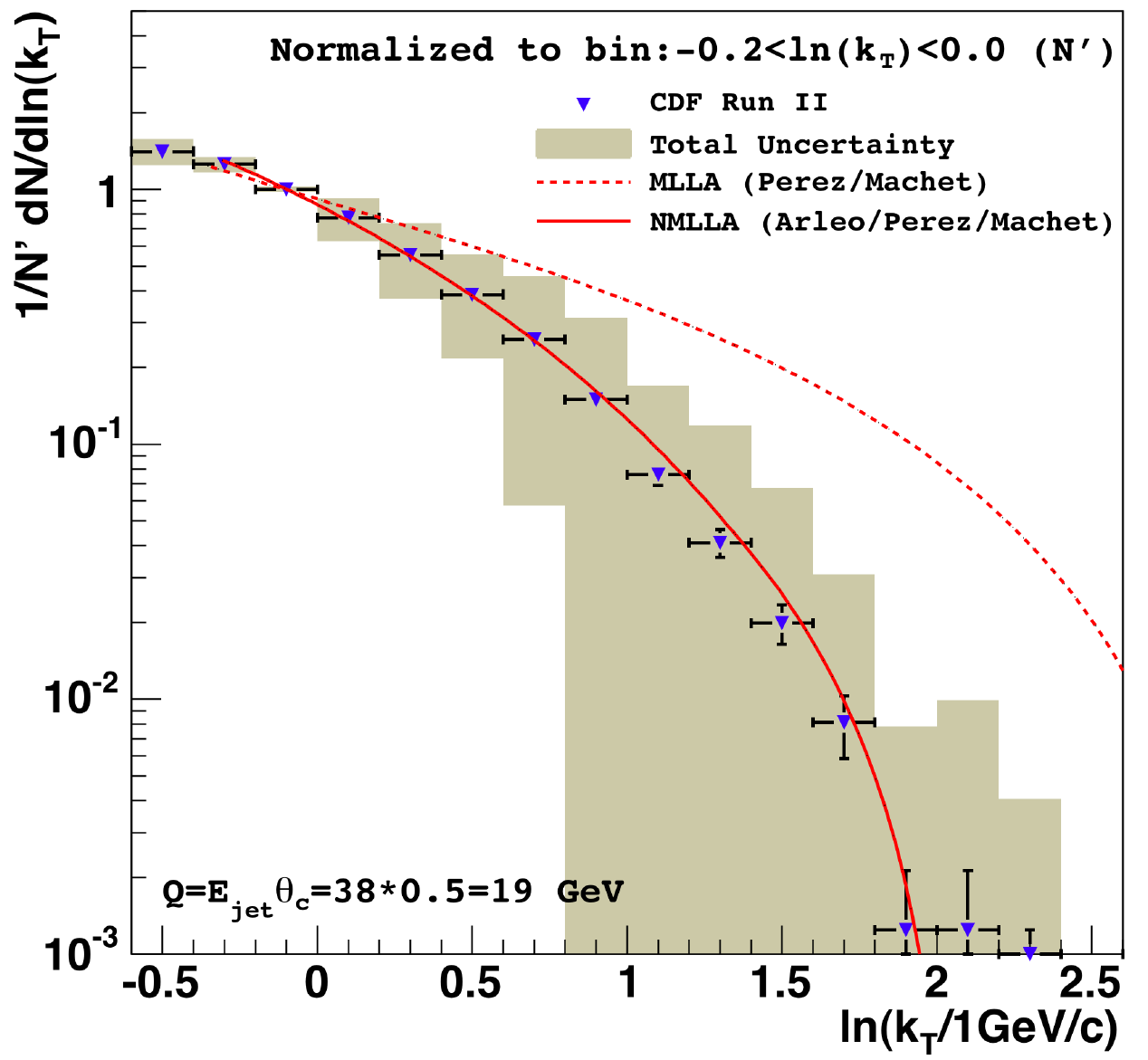}
\caption{[left] CDF central exclusive $\mu^+\mu^-$ invariant mass distribution for $p\bar{p}\to p\left[\mu^+\mu^-(\gamma)\right]\bar{p}$ processes~\cite{CDF:ExclDijetccbar}; [right] CDF $k_T$ distribution of particles in a cone of 0.5~rad around the jet axis in a dijet invariant mass bin with $\left<Q\right> = 19$~GeV, corresponding to dijet masses in the range $66 < M_{jj} < 95$~GeV/$c^2$~\cite{CDF:kTDist}.}
\label{fig:ccbar_kT}
\end{figure}

\vspace*{-0.3cm}
Central exclusive processes are those where the colliding hadrons emerge intact but impart $\gamma / g$ combinations that interact at higher order to produce fully measurable states at low rapidities, with surrounding rapidity regions devoid of particles.  CDF recently reported observations of $p \bar{p} \to p [{\rm dijet}] \bar{p}$, in dijet events with $E_T({\rm jet}) > 10$~GeV; and $p \bar{p} \to p \left[ \mu^+\mu^-(\gamma),\, J/\psi,\, \psi(2S),\, \chi_c^0\right]  \bar{p}$, with two oppositely charged central muons and either no other particles or one additional photon detected (see Fig.~\ref{fig:ccbar_kT})~\cite{CDF:ExclDijetccbar}.  Consideration of the SM Higgs $H\to b\bar{b}$ signature and quantum-number analogies between the scalar $\chi_c^0$ meson and the H boson permits these measurements to demonstrate the viability of exclusive SM Higgs production through $pHp$ processes at the LHC~\cite{CDF:ExclDijetccbar}.

\section{Jet Fragmentation, Underlying Event, and Minimum Bias Studies}
\vspace*{-0.3cm}
Soft QCD interactions in hadron collisions, studied through charged-particle observables, are experimentally and theoretically challenging yet crucial to understanding high $p_T$ phenomena.  A CDF study of the transverse momenta ($k_T$) of intra-jet particles with respect to their jet axis has provided insight into which stage of jet formation principally determines jet characteristics~\cite{CDF:kTDist}.  Fig.~\ref{fig:ccbar_kT} shows a $k_T$ distribution resembling resummed pQCD predictions, indicating support for local parton-hadron duality and a de-emphasis of non-perturbative hadronization effects in jet formation.

The underlying event (UE) consists of the beam-beam remnants minus the hard-scattering products and is becoming increasingly important to the discovery and precision potential at hadron colliders.  CDF has conducted an extensive program of UE studies that exploit jet and Drell-Yan event activity topologies to maximize the sensitivity of UE observables~\cite{CDF:UE}.  Refs.~\cite{CDF:UE} contain several distributions of UE-sensitive observables, corrected to the particle level, that suggest the UE may be universal (independent of the hard process) and inform MC tuning and development.

\begin{figure}
\includegraphics[width=.29\textwidth]{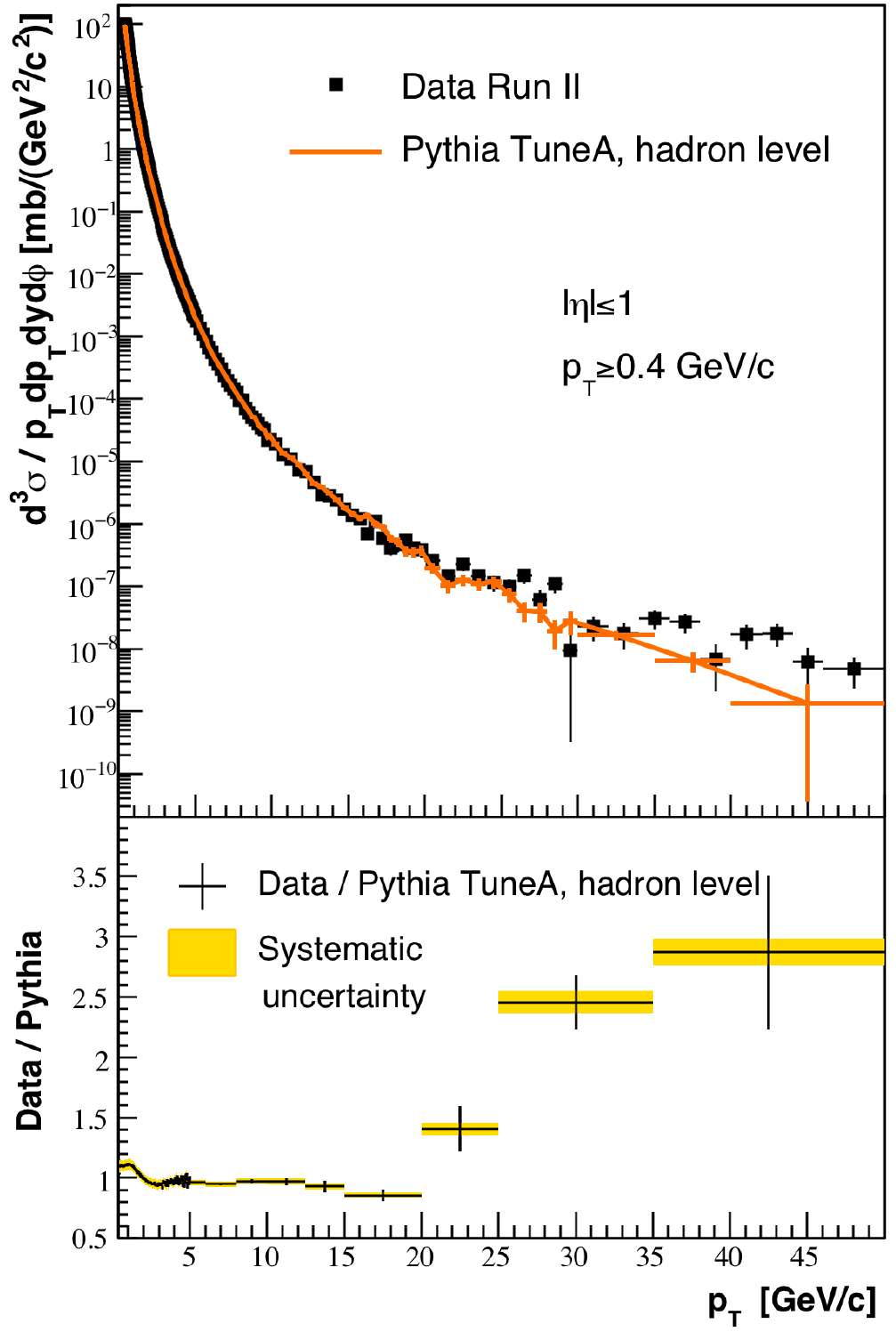}
\includegraphics[width=.36\textwidth]{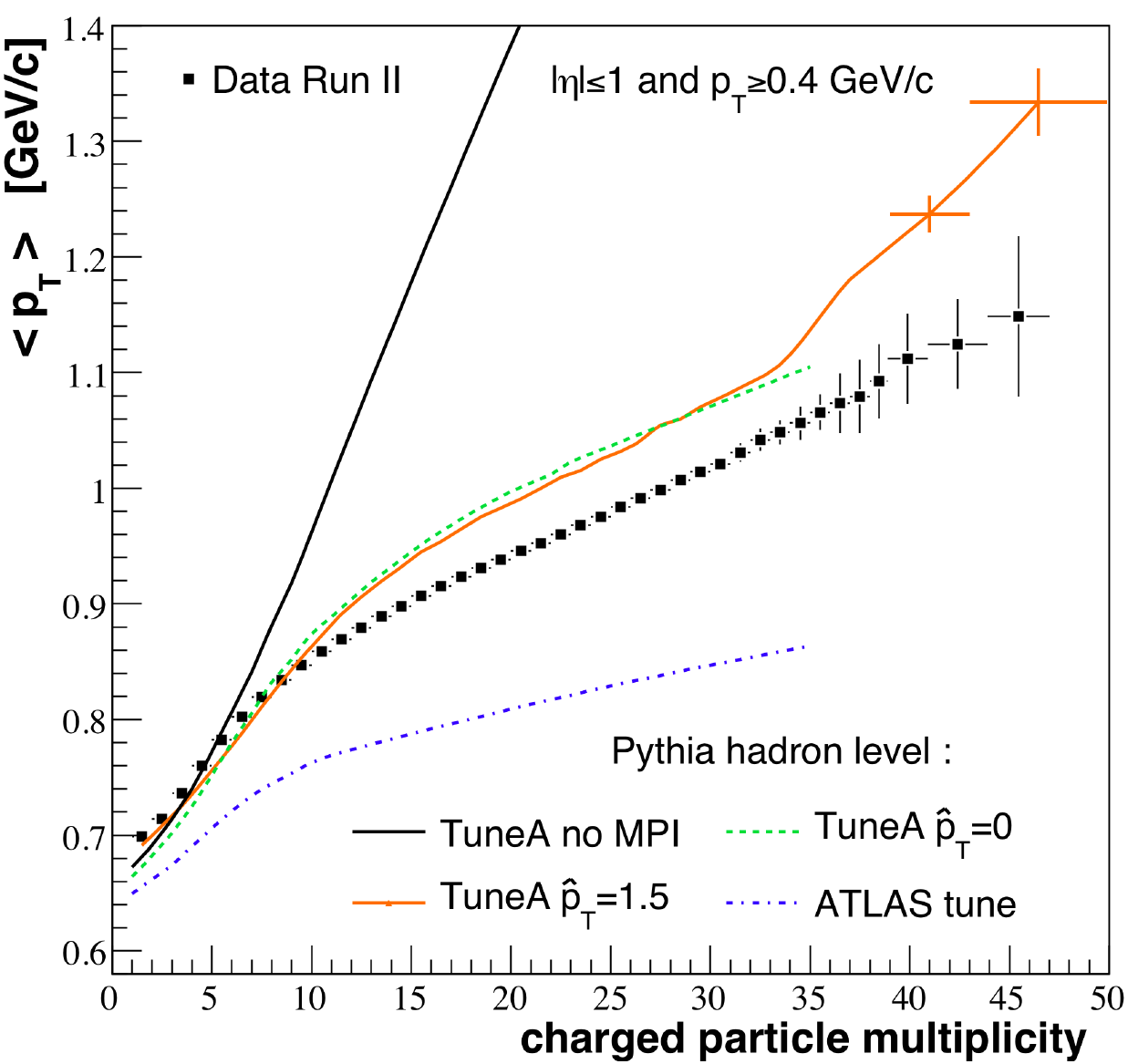}
\includegraphics[width=.31\textwidth]{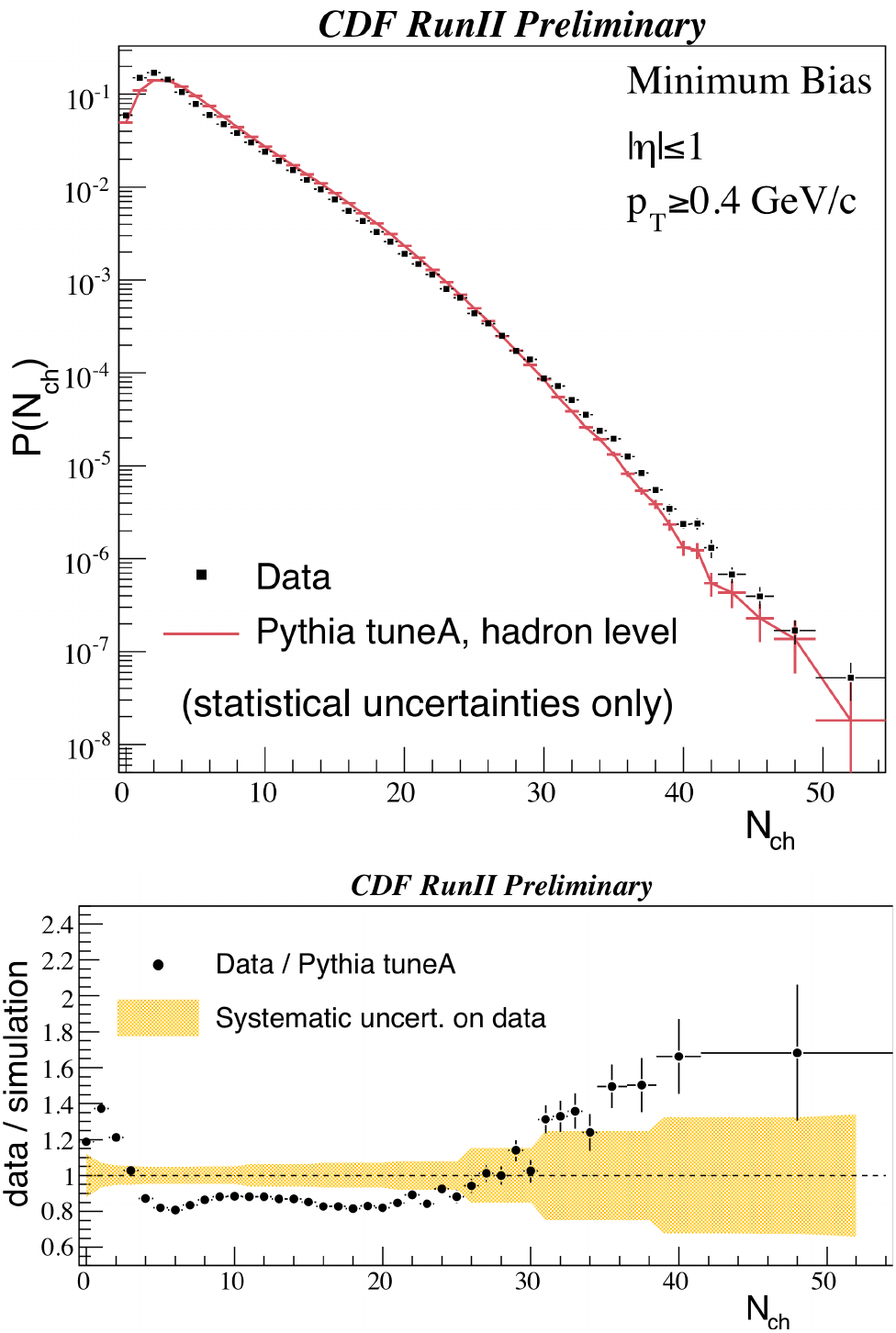}
\caption{CDF~\cite{CDF:minbias} track $p_T$ differential cross section [left], average track $p_T$ {\it vs.} multiplicity [centre] , and charged-particle multiplicity [right] distributions compared with various theoretical predictions.}
\label{fig:minbias}
\end{figure}

Other important inputs into MC tuning arise in inclusive inelastic $p\bar{p}$ collisions, studied using a minimum-bias trigger under low luminosity conditions.  CDF has measured an inclusive differential charged-particle $p_T$ cross section over 11 orders of magnitude (see Fig.~\ref{fig:minbias}), finding poor agreement with {\sc pythia} at higher momenta~\cite{CDF:minbias}.  Fig.~\ref{fig:minbias} also shows the measured correlation between the mean $p_T$ and multiplicity of charged particles in minimum-bias events, observables to which the MC tuning parameters favouring multiple parton interactions (MPI) are particularly sensitive.  The indicated charged-particle multiplicity distribution is the most precise and extensive ever measured in the $\left|\eta\right| \leq 1$ pseudorapidity range; existing MC calculations describe the high-multiplicity region poorly.

\vspace{-0.1cm}

\section{Concluding Remarks}
\vspace*{-0.3cm}
The Tevatron experiments now provide precision QCD physics at $\sqrt{s} = 1.96$~TeV, and measurements common to both CDF and D\O \ are mutually consistent.  The Rutherford scattering approach to studying the fundamental constituents of matter has now soundly entered the TeV regime, with stringent constraints on new physics, tests of pQCD, precise measurements of $\alpha_s$, insight into jet fragmentation processes, and information on high-$x$ PDF gluon content.  Theoretical improvements are called for in several places, in particular to describe the observed photon cross section and inclusive particle production event characteristics.  These results are based on less than one third of the anticipated complete Run~II sample, with more measurements expected in the coming years to illuminate the Large Hadron Collider era of QCD rediscovery and discovery, now begun!



\newpage

\vspace*{-1.0cm}

\begin{thebibliography}{99}

\bibitem{CDF:InclJet} T.~Aaltonen {\it et al.} (CDF), {\it Measurement of the inclusive jet cross section at the Fermilab Tevatron $p\bar{p}$ collider using a cone-based jet algorithm}, Phys.~Rev.~D {\bf 78}, 052006 (2008); {\it ibid.} {\bf 79}, 119902(E) (2009) [{\tt arXiv:0807.2204v4}].

\bibitem{D0:InclJet} V. M.~Abazov {\it et al.} (D\O ), {\it Measurement of the Inclusive Jet Cross Section in $p\bar{p}$ collisions at $\sqrt{s}$ = 1.96\ TeV}, Phys.~Rev.~Lett. {\bf 101}, 062001 (2008) [{\tt arXiv:0802.2400v3}].

\bibitem{D0:alpha_s} V. M.~Abazov {\it et al.} (D\O ), {\it Determination of the strong coupling constant from the inclusive jet cross section in $p\bar{p}$ collisions at $\sqrt{s}$ = 1.96\ TeV}, Phys.~Rev.~D {\bf 80}, 111107(R) (2009) [{\tt arXiv:0911.2710v3}].

\bibitem{CDF:alpha_s} T.~Affolder {\it et al.} (CDF), {\it Measurement of the Strong Coupling Constant from Inclusive Jet Production at the Tevatron $p\bar{p}$ Collider}, Phys.~Rev.~Lett. {\bf 88}, 042001 (2002) [{\tt arXiv:hep-ex/0108034v1}].

\bibitem{CDF:dijet_mass} T.~Aaltonen {\it et al.} (CDF), {\it Search for new particles decaying into dijets in proton-antiproton collisions at $\sqrt{s}$ = 1.96\ TeV}, Phys.~Rev.~D {\bf 79}, 112002 (2009) [{\tt arXiv:0812.4036v1}].

\bibitem{D0:dijet_mass} V. M.~Abazov {\it et al.} (D\O ), {\it Measurement of the Dijet Mass Cross Section in $p\bar{p}$ collisions at $\sqrt{s}$ = 1.96\ TeV}, D\O \ Note 5919-CONF, Apr.~29, 2009.

\bibitem{D0:dijet_angular} V. M.~Abazov {\it et al.} (D\O ), {\it Measurement of Dijet Angular Distributions at $\sqrt{s}$ = 1.96\ TeV and Searches for Quark Compositeness and Extra Spatial Dimensions}, Phys.~Rev.~Lett. {\bf 103}, 191803 (2009) [{\tt arXiv:0906.4819v2}].

\bibitem{CDF:photon} T.~Aaltonen {\it et al.} (CDF), {\it Measurement of the inclusive isolated prompt photon cross section in $p\bar{p}$ collisions at $\sqrt{s}$ = 1.96\ TeV using the CDF detector}, Phys.~Rev.~D {\bf 80}, 111106(R) (2009) [{\tt arXiv:0910.3623v2}].

\bibitem{D0:photon} V. M.~Abazov {\it et al.} (D\O ), {\it Measurement of the isolated photon cross section in $p\bar{p}$ collisions at $\sqrt{s}$ = 1.96\ TeV}, Phys.~Lett.~B {\bf 639}, 151 (2006); {\it ibid.} {\bf 658}, 285(E) (2008) [{\tt arXiv:hep-ex/0511054v1}].

\bibitem{D0:DPS} V. M.~Abazov {\it et al.} (D\O ), {\it Double parton interactions in photon + 3 jet events in $p\bar{p}$ collisions at $\sqrt{s}$ = 1.96\ TeV}, {\tt arXiv:0912.5104v1}.

\bibitem{CDF:DPS} F.~Abe {\it et al.} (CDF), {\it Double parton scattering in $\bar{p}p$ collisions at $\sqrt{s}$ = 1.8\ TeV}, Phys.~Rev.~D {\bf 56}, 3811 (1997).

\bibitem{Snigirev} A. M.~Snigirev, {\it A possible indication to the QCD evolution of double parton distributions?} {\tt arXiv:1001.0104v2}.

\bibitem{CDF:ExclDijetccbar} T.~Aaltonen {\it et al.} (CDF), {\it Observation of exclusive dijet production at the Fermilab Tevatron $p\bar{p}$ collider}, Phys.~Rev.~D {\bf 77}, 052004 (2008) [{\tt arXiv:0712.0604v3}]; {\it idem}, {\it Observation of Exclusive Charmonium Production and $\gamma\gamma\to\mu^+\mu^-$ in $p\bar{p}$ Collisions at $\sqrt{s}$ = 1.96\ TeV}, Phys.~Rev.~Lett. {\bf 102}, 242001 (2009) [{\tt arXiv:0902.1271v4}].

\bibitem{CDF:kTDist} T.~Aaltonen {\it et al.} (CDF), {\it Measurement of the $k_T$ Distribution of Particles in Jets Produced in $p\bar{p}$ Collisions at $\sqrt{s}$ = 1.96\ TeV}, Phys.~Rev.~Lett. {\bf 102}, 232002 (2009) [{\tt arXiv:0811.2820v1}].

\bibitem{CDF:UE} T.~Affolder {\it et al.} (CDF), {\it Charged jet evolution and the underlying event in proton-antiproton collisions at 1.8 TeV}, Phys.~Rev.~D {\bf 65}, 092002 (2002); D.~Acosta {\it et al.} (CDF), {\it Underlying event in hard interactions at the Fermilab Tevatron $p\bar{p}$ collider}, Phys.~Rev.~D {\bf 70}, 072002 (2004) [{\tt arXiv:hep-ex/0404004v1}]; D.~Kar {\it et al.} (CDF), {\it Measurement of the Underlying Event at Tevatron}, {\tt arXiv:0905.2323v1}.

\bibitem{CDF:minbias} T.~Aaltonen {\it et al.} (CDF), {\it Measurement of particle production and inclusive differential cross sections in $p\bar{p}$ Collisions at $\sqrt{s}$ = 1.96\ TeV}, Phys.~Rev.~D {\bf 79}, 112005 (2009) [{\tt arXiv:0904.1098v2}]; {\it idem}, {\it Multiplicity Distribution of Charged Particles in Inelastic $p\bar{p}$ Interactions}, CDF-9936, Sept.~22, 2009.

\end{thebibliography}
\end{document}